\documentclass[aip,jcp,twocolumn,superscriptaddress]{revtex4}%
\usepackage{epsfig,amssymb,amsmath,amsthm,amsfonts,amsbsy,mathrsfs}
\usepackage{graphicx,color}
\usepackage{amsmath}
\usepackage{amssymb}%
\usepackage{epstopdf}
\begin{document}
\title{Strong Edge Magnetism and Tunable Energy Gaps in Zigzag Graphene
Quantum Dots with High Stability}

\author{Wei Hu}
\affiliation{Computational Research Division, Lawrence Berkeley
National Laboratory, Berkeley, California 94720, United States}

\author{Yi Huang}
\affiliation{Department of Applied Physics, Xi'an Jiaotong
University, Xi'an, Shaanxi 710049, China}

\author{Lin Lin}
\affiliation{Department of Mathematics, University of California,
Berkeley, California 94720, United States}
\affiliation{Computational Research Division, Lawrence Berkeley
National Laboratory, Berkeley, California 94720, United States}

\author{Erjun Kan}
\affiliation{Department of Applied Physics and Institution of Energy
and Microstructure, Nanjing University of Science and Technology,
Nanjing, Jiangsu 210094, China}

\author{Xingxing Li}
\affiliation{Hefei National Laboratory for Physical Sciences at
Microscale, Department of Chemical Physics, and Synergetic
Innovation Center of Quantum Information and Quantum Physics,
University of Science and Technology of China, Hefei, Anhui 230026,
China}

\author{Chao Yang}
\affiliation{Computational Research Division, Lawrence Berkeley
National Laboratory, Berkeley, California 94720, United States}

\author{Jinlong Yang}
\affiliation{Hefei National Laboratory for Physical Sciences at
Microscale, Department of Chemical Physics, and Synergetic
Innovation Center of Quantum Information and Quantum Physics,
University of Science and Technology of China, Hefei, Anhui 230026,
China}

\date{\today}

\pacs{ }

\begin{abstract}

Graphene is a nonmagnetic semimetal and cannot be directly used as
electronic or spintronic devices. We demonstrate that graphene
quantum dots (GQDs) can exhibit strong edge magnetism and tunable
energy gaps due to the presence of localized edge states. By using
large-scale first principle density functional theory (DFT)
calculations and detailed analysis based on model Hamiltonians, we
can show that the zigzag edge states in GQDs become much stronger
and more localized as the system size increases. The enhanced edge
states induce strong electron-electron interactions along the edges
of GQDs, ultimately resulting in a magnetic phase transition from
nonmagnetic to intra-edge ferromagnetic and inter-edge
antiferromagnetic, when the diameter is larger than 4.5 nm. Our
analysis shows that the inter-edge superexchange interaction of
antiferromagnetic states between two nearest-neighbor zigzag edges
in GQDs is much stronger than that exists between two parallel
zigzag edges in GQDs and graphene nanoribbons. Furthermore, such
strong and localized edge states also induce GQDs semiconducting
with tunable energy gaps, mainly controlled by adjusting the system
size. Our results show that the quantum confinement effect,
inter-edge superexchange (antiferromagnetic), and intra-edge direct
exchange (ferromagnetic) interactions are crucial for the electronic
and magnetic properties of zigzag GQDs at the nanoscale.

\end{abstract}

\maketitle


Engineering techniques that use finite size effect to introduce
tunable edge magnetism and energy gap are by far the most promising
ways for enabling graphene~\cite{Scinece_306_666_2004} to be used in
electronics and spintronics~\cite{NatureMater_6_183_2007,
RMP_81_109_2009}. Examples of finize sized graphene nanostructures
include one-dimensional (1D) graphene nanoribbons
(GNRs)~\cite{PRL_98_206805_2007, Science_319_1229_2008,
Science_323_1701_2009, Nature_444_347_2006, PRL_97_216803_2006,
PRL_99_186801_2007, JACS_130_4224_2008, JACS_131_17728_2009,
PRL_102_227205_2009, Nature_514_608_2014} and zero-dimensional (0D)
graphene nanoflakes (GNFs) (also known as graphene quantum dots
(GQDs))~\cite{Science_320_356_2008, AdvFunctMater_18_3506_2008,
PRL_100_056807_2008, NatureMater_8_235_2009, PRB_81_085430_2010,
PRB_82_045409_2010, AdvMater_22_505_2010, PCCP_19_6338_2017,
JPCC_116_5531_2012, Carbon_67_721_2014, JCP_140_074304_2014,
JCP_141_214704_2014_GNFs}. It is well known that electronic and
magnetic properties~\cite{RepProgPhys_73_056501_2010} of GNRs and
GNFs depend strongly on the atomic configuration of their edges,
which are of either the armchair (AC) or zigzag (ZZ)
types~\cite{PRL_97_216803_2006}.

Edge magnetism has been predicted
theoretically~\cite{JACS_130_4224_2008} and observed
experimentally~\cite{Nature_514_608_2014} in ZZGNRs. The magnetism
results from the antiferromagnetic (AFM) coupling between two
parallel ferromagnetic (FM) zigzag edges of ZZGNRs. However, the
inter-edge superexchange interaction of such AFM states in ZZGNRs
rapidly weakens ($\sim$ $w^{-2}$) as the ribbon-width $w$
increases~\cite{PRL_102_227205_2009}. Furthermore, the energy gap of
GNRs depend on several factors, such as the edge type (armchair or
zigzag) and the width of the nanoribbon~\cite{PRL_97_216803_2006},
thus cannot be easily tuned. Such problem does not exist in GNFs due
to the quantum confinement effect~\cite{PRL_90_037401_2003}. The
ability to the control GNF energy gap has enabled GNFs to be used in
promising applications in electronics~\cite{NatureMater_8_235_2009}.
In addition, triangular ZZGNFs are theoretically predicted to have
strong edge magnetism even in small
systems~\cite{PRL_99_177204_2007, NanoLett_8_241_2008}. Recent
experiments~\cite{npjQuantumMaterials_2_5_2017} have also
demonstrated that edge magnetism can be observed in ZZGNFs when the
edges are passivated by certain chemical groups. However, triangular
ZZGNFs have large formation energy~\cite{PCCP_19_6338_2017} and have
not been synthesized experimentally. Interestingly, hexagonal ZZGNFs
exhibits significantly improved stability in ambient
environment~\cite{PCCP_19_6338_2017}.  Theoretically, semi-empirical
tight-binding model~\cite{PRB_77_235411_2008, PRB_82_155445_2010}
and first principle density functional theory (DFT)
calculations~\cite{JPCC_116_5531_2012, Carbon_67_721_2014,
JCP_140_074304_2014, JCP_141_214704_2014_GNFs} for hexagonal ZZGNFs
have been performed for small sized systems but found no magnetism
(NM). Thus the prospect of finding stable finite sized graphene
easily fabricated in experiments that exhibits both strong edge
magnetism and tunable energy gap seems dim.

In this letter, we systematically study the electronic and magnetic
properties of hexagonal ZZGNFs with the diameters in the range of 2
nm to 12 nm (with up to 3900 atoms). Using first-principles DFT
calculations, we find that both strong edge magnetism and tunable
energy gap can be realized simultaneously in large ZZGNFs. We
demonstrate that spin polarization plays a crucial role as the
diameter of a ZZGNF increases beyond 4.5 nm. A spin-unpolarized
calculation shows that edge states become increasingly more
localized as the size a ZZGNF increases. These edge states form a
half-filled pseudo-band and is thus unstable. Adding
spin-polarization allows the edge states to spontaneously split into
spin-polarized occupied and unoccupied states. This separation
results in a magnetic phase transition from an NM phase to a strong
inter-edge AFM phase. It also opens a tunable band gap that can be
easily controlled by quantum confinement effect. These properties
make GNFs better candidate materials for nanoelectronics than
GNRs~\cite{PRL_97_216803_2006}. We also confirm that ZZGNFs
passivated by different chemical groups all exhibit similar
behavior. Such flexibility may facilitate future experimental
synthesis of such ZZGNFs.


We use the Kohn-Sham DFT based electronic structure analysis tools
implemented in the SIESTA (Spanish Initiative for Electronic
Simulations with Thousands of Atoms)~\cite{JPCM_14_2745_2002_SIESTA}
software package. We use the generalized gradient approximation
of Perdew, Burke, and Ernzerhof
(GGA-PBE)~\cite{PRL_77_3865_1996_PBE} exchange correlation
functional with collinear spin polarization, and
the double zeta plus polarization orbital basis set (DZP) to
describe the valence electrons within the framework of a linear
combination of numerical atomic orbitals
(LCAO)~\cite{PRB_64_235111_2001_LCAO}. All atomic coordinates are
fully relaxed using the conjugate gradient (CG) algorithm until the
energy and force convergence criteria of 10$^{-4}$ eV and 0.02
eV/{\AA} respectively are reached. Due to the large number of atoms
contained in hexagonal hydrogen-passivated ZZGNFs
(C$_{6n^2}$H$_{6n}$, $n$ = 1 $\sim$ 25 ), we use the recently
developed PEXSI (Pole EXpansion and Selected Inversion)
method~\cite{LinLuYingEtAl2009, JPCM_25_295501_2013_PEXSI,
JPCM_26_305503_2014_PEXSI} to accelerate the computation.

\begin{figure}[htbp]
\begin{center}
\includegraphics[width=0.5\textwidth]{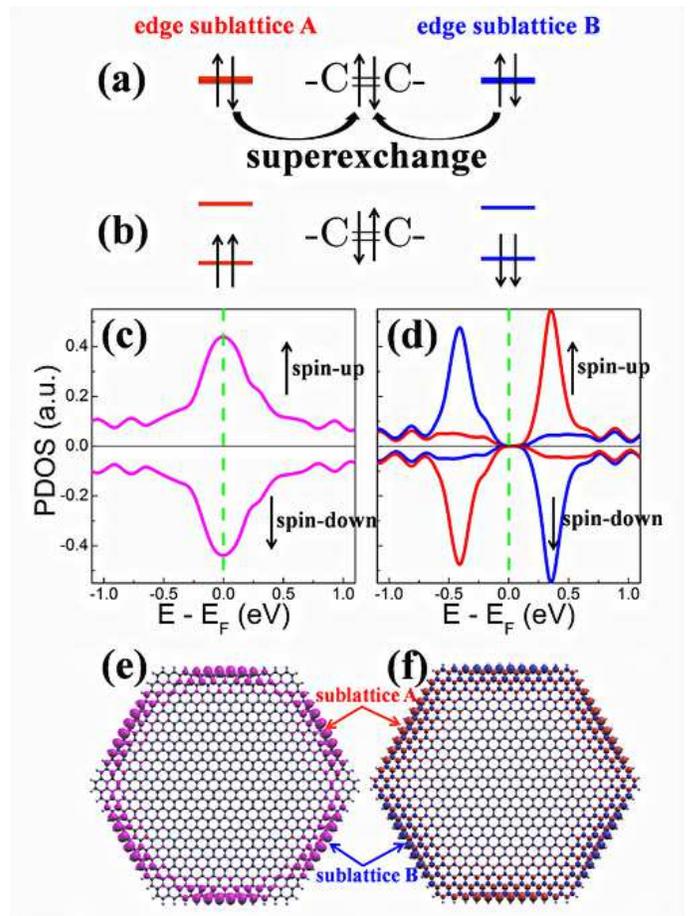}
\end{center}
\caption{Electronic structure of edge states in C$_{864}$H$_{72}$ in
two different magnetic phases (NM and AFM), including the schematic
illustration of orbital diagram of superexchange interaction of edge
states in the (a) NM and (b) AFM phases, projected density of states
(PDOS) of edges in the (c) NM and (d) AFM phases, (e) local density
of states (LDOS) of Fermi level (pink isosurfaces) in the NM phase
and (f) spin density isosurfaces in the AFM phase. The red and blue
isosuraces in (f) represent the spin-up and spin-down states,
respectively. The red and blue lines in (d) represent the PDOS
contributed by sublattice A (spin-up edges) and B (spin-down edges)
atoms in graphene, respectively. The fermi level is marked by green
dotted lines and set to zero.} \label{fig:C864H72}
\end{figure}

We demonstrate the importance of spin polarization using
C$_{864}$H$_{72}$ (6 nm) as an example (Figure~\ref{fig:C864H72}).
In spin unpolarized calculations, strong and localized edge states
are observed (Figure~\ref{fig:C864H72}(e)), which induce high
electron density on the edges of C$_{864}$H$_{72}$. Furthermore,
such edge states in ZZGNFs becomes much stronger and more localized
as the sizes increase~\cite{JCP_141_214704_2014_GNFs}, which lead to
metallic ZZGNFs at the nanoscale~\cite{PRB_82_155445_2010}.
Figure~\ref{fig:C864H72}(c) plots the projected density of states
(PDOS) of the carbon edges of C$_{864}$H$_{72}$ and shows
considerable high density of states (DOS) near the Fermi level. This
confirms that C$_{864}$H$_{72}$ exhibits metallic characteristics in
spin unpolarized calculations due to the strong localized edge
states~\cite{JCP_141_214704_2014_GNFs}.


In spin polarized calculations, such half-filled metallic edge
states are not stable, and can spontaneously split into two types of
occupied and unoccupied states as shown in
Figure~\ref{fig:C864H72}(b) and (d). A magnetic phase transition
occurs from the NM phase to a magnetic phase that exhibits
intra-edge FM and inter-edge AFM characters as shown in
Figure~\ref{fig:C864H72}(f). This can be interpreted as the
Mott-type competition between the kinetic (hopping) energy and the
intra-edge (on-site) electron-electron interaction energy with
respect to the spin polarized edge states.
The minimization of the kinetic energy tends to produce delocalized
spin states across all edges, while the minimization of the
electron-electron interaction energy tends to penalize simultaneous
occupation of the same edge by spin up and spin down electrons. Our
calculation indicates that for small system sizes, the kinetic
energy dominates, which agrees with previous theoretical prediction
of the NM phase for hexagonal ZZGNFs~\cite{PRB_77_235411_2008,
PRB_82_155445_2010,JPCC_116_5531_2012, Carbon_67_721_2014,
JCP_140_074304_2014, JCP_141_214704_2014_GNFs}. Only as the system
size increases, the effective electron-electron interaction energy
for the edge states starts to dominate and results in the phase
transition.

\begin{figure}[htbp]
\begin{center}
\includegraphics[width=0.5\textwidth]{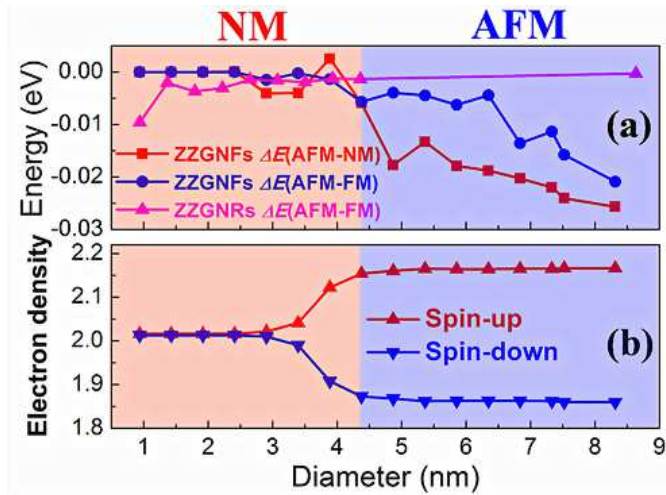}
\end{center}
\caption{(a) Relative energy per edge atom ($\Delta$$E$(AFM-NM) and
$\Delta$$E$(AFM-FM)) of NM, AFM and FM coupling between different
edges in ZZGNFs and ZZGNRs and (b) spin electron density
$<\hat{n}_{i{\sigma}}>$ (${\sigma}$ = ${\uparrow}$ (spin-up) or
${\downarrow}$ (spin-down)) at the carbon atom $i$ in the middle of
each zigzag edge in AFM ZZGNFs under the variation of the diameter
(ZZGNFs) or ribbon-with length (ZZGNRs). The red and blue regions
represent the stable NM ($\Delta$$E$(AFM-NM) $\approx$ 0) and AFM
($\Delta$$E$(AFM-NM) $<$ 0) coupling between different edges in
ZZGNFs, respectively.} \label{fig:Ef}
\end{figure}

Figure~\ref{fig:Ef}(a) shows the variation of relative energy of NM,
AFM and FM coupling between different edges in ZZGNFs and ZZGNRs
with respect to the system size. Our calculations show the AFM
states are much more stable than the NM and FM states in large-scale
cases, and a magnetic phase transition (Figure~\ref{fig:Ef}(b))
occurs in ZZGNFs as the diameter increases larger than 4.5 nm
(C$_{486}$H$_{54}$)~\cite{PRL_99_177204_2007}. In detail, the
intra-edge direct exchange interactions induce FM states along each
zigzag edge (belong to the same sublattice) and the inter-edge
superexchange interactions induce AFM states between two
nearest-neighbor edges (belong to different sublattices) through a
carbon-carbon double bond (C=C) at the corner in ZZGNFs
(Figure~\ref{fig:C864H72}(a)) at the nanoscale. The local magnetic
moment $M_i$ = $\mid<\hat{n}_{i{\uparrow}}>$ -
$<\hat{n}_{i{\downarrow}}>\mid$ ($<\hat{n}_{i{\sigma}}>$ is spin
electron density and ${\sigma}$ = ${\uparrow}$ (spin-up) or
${\downarrow}$ (spin-down)) at the carbon atom $i$ (defined to be
the one with the largest magnetic moment in the middle of each
zigzag edge in ZZGNFs) increases with the system size, and converges
to 0.3 $uB$ when the diameter is larger than than 6 nm
(C$_{864}$H$_{72}$). Furthermore, there is no charge transfer
($<\hat{n}_{i{\uparrow}}>$ + $<\hat{n}_{i{\downarrow}}>$ $\approx$
4) between different edge carbon atoms (belong to the same or
different sublattices) in ZZGNFs as the system size increases.

Notice that the intra-edge direct exchange interactions of FM sates
along each zigzag edge in ZZGNFs are similar to that in ZZGNRs.
However, the inter-edge superexchange interactions of AFM states
between two nearest-neighbor edges through a C=C bond at the corner
in ZZGNFs (Figure~\ref{fig:C864H72}(a)) are much stronger than that
between two parallel edges through $\pi$-electron in ZZGNRs as shown
in Figure~\ref{fig:Ef}(a), where such AFM coupling weakens rapidly
as the ribbon-width increases~\cite{PRL_102_227205_2009}. Our DFT
calculations confirm that the energy difference of AFM and FM
coupling between two parallel edges in large-scale 1D ZZGNRs is
negligible compared to ZZGNFs reported here.

\begin{figure}[htbp]
\begin{center}
\includegraphics[width=0.5\textwidth]{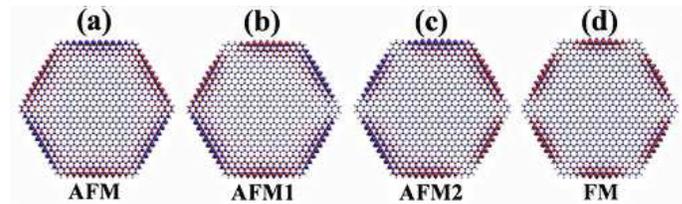}
\end{center}
\caption{Spin density isosurfaces of hydrogen-passivated
C$_{864}$H$_{72}$ in four different magnetic states, three types of
antiferromagnetic (AFM, AFM1 and AFM2) and one type of ferromagnetic
(FM) coupling at the inter edges. The red and blue isosurfaces
represent the spin-up and spin-down states, respectively.}
\label{fig:C864H72AFMFM}
\end{figure}

The enhanced stability of spin-polarized ZZGNFs can be understood by
using the Heisenberg model. We consider each FM edge as one site and
count the edge magnetic exchange interactions, and the Hamiltonian
can be written as
\begin{equation}
\hat{H} = - \sum J_{i,j}\vec{M_i}\vec{M_j} \label{eq:Heisenberg}
\end{equation}
where $J_{i,j}$ is the exchange parameter between two states $i$ and
$j$, $\vec{M_i}$ and $\vec{M_j}$ are corresponding spin magnetic
moments (The details are given in the Supplemental Material). There
are four different magnetic states in C$_{864}$H$_{72}$, three types
of antiferromagnetic (AFM, AFM1 and AFM2) and one type of
ferromagnetic (FM) coupling at the edges as shown in
Figure~\ref{fig:C864H72AFMFM}. The total energies of magnetic phases
$E(\text{AFM})$, $E(\text{AFM1})$, $E(\text{AFM2})$ and
$E(\text{FM})$ can be computed by the DFT calculations, and the the
exchange parameters can be evaluated by~\cite{JACS_136_5664_2014}
\begin{equation}
\begin{split}
E(\text{AFM}) &= (6J_1 - 6J_2 + 3J_3)M^2 + E_0  \\
E(\text{AFM1}) &= (2J_1 + 2J_2 - J_3)M^2 + E_0  \\
E(\text{AFM2}) &= (-J_1 + 2J_2 + 3J_3)M^2 + E_0  \\
E(\text{FM}) &= (-6J_1 - 6J_2 - 3J_3)M^2 + E_0
\end{split}
\end{equation}
where $J_1$, $J_2$ and $J_3$ are ortho-, meta- and para- edge
exchange interaction parameters, respectively. $M$ is the spin
magnetic moment at each edge. $E_0$ is nonmagnetic reference total
energy. We find that inter-edge exchange strength ($J_1$ = -0.038351
eV, $J_2$ = 0.000954 eV and $J_3$ = 0.001633 eV for
C$_{864}$H$_{72}$) between two nearest-neighbor edges is ten times
stronger than that between two parallel edges in ZZGNFs and ZZGNRs.
Therefore, ZZGNFs can maintain strong edge magnetism as the system
size increases, superior to that in
ZZGNRs~\cite{Nature_444_347_2006, PRL_102_227205_2009}.

We also check the effects of different types of passivating atoms
(bare and fluorine) and the shape (non-hexagonal) of ZZGNFs on their
electronic and magnetic properties, and find that the results of
magnetic phase transition and semiconductor characteristics are
similar to that in hexagonal hydrogen-passivated ZZGNFs (The details
are given in the Supplemental Material)~\cite{PRB_90_035403_2014}.
Such robustness provides flexibility in terms of the chemical
environment of ZZGNFs, and thus may facilitate the synthesis of
large scales ZZGNFs with tunable edge magnetism and energy gaps as
candidates for electronic and spintronic
devices~\cite{PhysStatusSolidiRRL_10_11_2016}.


We remark that the magnetic phase transition and associated with
tunable electronic structures, especially energy gaps, can also be
observed in the Hubbard model~\cite{PRL_99_177204_2007}. From our
first principle calculations, we find that choosing the parameters
$t$ = 2.5 eV and $U$ = 2.1 eV in the Hubbard model can well
reproduce the size-dependent energy gaps (The details are given in
the Supplemental Material). Figure~\ref{fig:Eg} plots how the
HOMO-LUMO energy gap $E_g$ change with respect to the system size of
ZZGNFs and ACGNFs in two different magnetic phases (NM and AFM). Our
DFT calculations and mean-field Hubbard model show similar results
that the energy gaps $E_g$ of ZZGNFs decrease as the system size
increases. In particular, we find that the energy gap of NM ZZGNFs
decreases more rapidly with respect to the system size than that in
AFM ZZGNFs, due to the presence of edge states whose electron
density near the edges of ZZGNFs as shown in
Figure~\ref{fig:C864H72}(e). This observation is consistent with
previous results obtained from tight-binding
models~\cite{PRB_77_235411_2008, PRB_82_155445_2010} and DFT
calculations~\cite{JCP_140_074304_2014, JCP_141_214704_2014_GNFs}.
However, AFM semiconducting ZZGNFs show similar scaling behavior of
the energy gap to NM ACGNFs at the
nanoscale~\cite{JCP_141_214704_2014_GNFs}. Therefore, edge states
should have little effect on the energy gaps of AFM ZZGNFs and the
quantum confinement effect\cite{PRL_90_037401_2003} is the only
factor to control the energy gaps in ZZGNFs and ACGNFs
(Figure~\ref{fig:Eg}(a)). In detail, NM ZZGNFs exhibits metallic
characters ($E_g$ is smaller than the thermal fluctuation (25 $meV$)
at room temperature) when the diameter is larger than 7 nm
(C$_{1350}$H$_{90}$), but AFM ZZGNFs with the diameter of 12 nm
(C$_{3750}$H$_{150}$) still behaves as a semiconductor with a
sizable energy gap $E_g$ = 0.23 eV, similar to the case of NM
ACGNFs~\cite{JCP_141_214704_2014_GNFs}.
\begin{figure}[htbp]
\begin{center}
\includegraphics[width=0.5\textwidth]{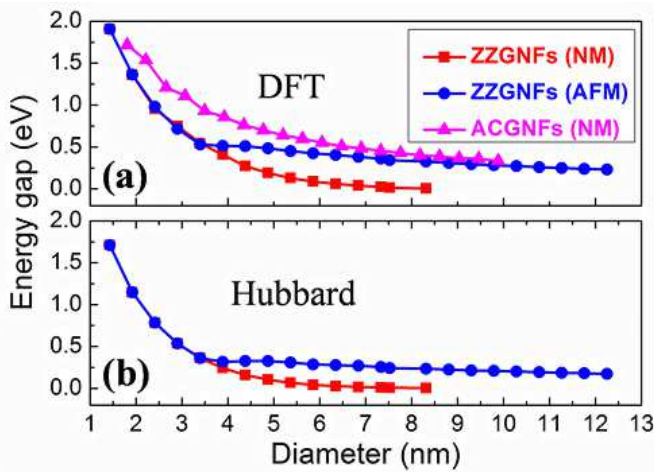}
\end{center}
\caption{Energy gap $E_g$ (eV) of ZZGNFs and ACGNFs in two different
magnetic phases (NM and AFM) as a function of the diameter size
(nm), computed with two different methods, (a) DFT calculations and
(b) Hubbard model ($t$ = 2.5 eV and $U$ = 2.1 eV).} \label{fig:Eg}
\end{figure}


In summary, using large scale first principle calculations, we show that
the electronic and magnetic properties of hexagonal zigzag graphene
quantum dots (GQDs) can be significantly affected by the system size.
We found that the zigzag edge states in GQDs become much stronger and more
localized as the system size increases. The presence of these edge
states induce strong electron-electron interactions along the edges
of GQDs, resulting in a magnetic phase transition from nonmagnetic
to intra-edge ferromagnetic and inter-edge antiferromagnetic when
the diameter is larger than 4.5 nm. On the other hand, such strong
and localized edge states also induce GQDs semiconducting with
tunable energy gaps only controlled by adjusting the system size.
Therefore, ZZGNFs with strong edge magnetism and tunable energy gaps
may be promising candidates and practical electronic and spintronic
applications.  \\


This work was performed, in part, under the auspices of the U.S.
Department of Energy by Lawrence Livermore National Laboratory under
Contract DE-AC52-07NA27344. Support for this work was provided
through Scientific Discovery through Advanced Computing (SciDAC)
program funded by U.S.~Department of Energy, Office of Science,
Advanced Scientific Computing Research and Basic Energy Sciences (W.
H., L. L. and C. Y.), by the Center for Applied Mathematics for
Energy Research Applications (CAMERA), which is a partnership
between Basic Energy Sciences and Advanced Scientific Computing
Research at the U.S Department of Energy (L. L. and C. Y.), and by
the Department of Energy under Grant No. DE-SC0017867 (L. L.). This
work is also partially supported by the National Key Research and
Development Program of China (Grant No. 2016YFA0200604) and the
National Natural Science Foundation of China (NSFC) (Grant No.
21688102 and 51522206). Y. H. acknowledges support from the
Education Program for Talented Students of Xi'an Jiaotong
University. We thank the National Energy Research Scientific
Computing (NERSC) center, and the USTCSCC, SC-CAS, Tianjin, and
Shanghai Supercomputer Centers for the computational
resources.  \\

\footnotesize{

}

\end{document}